\documentclass[aps,prb,twocolumn,showpacs,superscriptaddress]{revtex4-1}\begin{tiny}\end{tiny}
\usepackage{longtable}
\usepackage{amsfonts}
\usepackage{graphicx}
\usepackage{setspace}

\begin{document}

\title{Electron Self-Energy and Generalized Drude Formula \\
for Infrared Conductivity of Metals}

\author{Philip B. Allen }
\affiliation{Department of Physics and Astronomy, 
State University of New York, Stony Brook, NY 11794-3800}

\date{\today}

\begin{abstract}
G\"otze and W\"olfle (GW) wrote the
conductivity in terms of a memory function
$M(\omega)$ as $\sigma(\omega+i\eta)=(ine^2/m)(\omega+M(\omega+i\eta))^{-1}$,
where $M(\omega+i\eta)=i/\tau$ in the Drude limit.
The analytic properties of $-M(\omega+i\eta)$ are the same as those of the
self-energy $\Sigma$ of a retarded Green's function.  In 
the approximate treatment of GW, $-M$ closely resembles a self-energy,
with differences, {\it e.g.}, the imaginary part is twice too large.
The correct relation between $-M$ and $\Sigma$ is
known for the electron-phonon case and is conjectured
to be similar for other perturbations.  When vertex
corrections are ignored there is a known relation.
A derivation using Matsubara temperature Green's functions is given.
\end{abstract}


\maketitle

\section{\label{sec:l}Preliminaries}

Holstein \cite{Holstein1}  used elementary arguments 
to show that in the infrared properties of metals
there can be quantum effects (outside of the semiclassical
Boltzmann approach) when the temperature is low enough and
the probing frequency $\omega$ 
is degenerate with non-electronic excitations like phonons.
Such effects have been seen experimentally \cite{Joyce,Farnworth}.
G\"otze and W\"olfle (GW) \cite{Gotze} gave a nice simplified
way to compute such effects in the optical response of metals using 
truncated equations of motion to compute the ``memory function''
$M(\omega+i\eta)$ defined as
\begin{equation}
\sigma(\omega+i\eta)=\frac{ine^2/m}{\omega+M(\omega+i\eta)}.
\end{equation}
In the dc limit, their formulas 
correctly reproduce lowest-order variational solutions of 
the corresponding Boltzmann transport theories.  
Unfortunately, a systematic perturbation theory for $M(\omega+i\eta)$
is not known, and the GW approximation
is therefore hard to improve.  The GW results are slightly less accurate than the
corresponding lowest-order results of Green's function
theories.

The function $-M(\omega+i\eta)$ has causal
analytic properties and, not surprisingly, bears a close resemblance to
an electron self-energy $\Sigma(\vec{k},\omega+i\eta)$ for $\vec{k}$-points
averaged over the Fermi surface.  However, the imaginary part of 
$\Sigma$ is $-1/2\tau$ while the imaginary part of $-M$ must
be $-1/\tau$.  This is not the only difference between $-M$ and $\Sigma$.
Since the analogy between $-M$ and $\Sigma$ is sometimes used
for analysis of infrared spectra \cite{Timusk}, it is important to 
understand just how good it actually is.  A full formula seems not to have
been derived, and is beyond the ambition
of this paper.  A reasonable conjecture is that when
anisotropy with $k$ around the Fermi surface is not too important, then
\begin{equation}
\sigma(\omega+i\eta)=\frac{ine^2}{m\omega} \int_{-\infty}^{\infty} d\omega^{\prime} 
\frac{f(\omega^{\prime})-f(\omega^{\prime}+\omega)}
{\omega-\Sigma_{\rm ir}(\omega^{\prime}+\omega+i\eta)
+\Sigma_{\rm ir}^{\ast}(\omega^{\prime}+i\eta)},
\label{eq:result}
\end{equation}
where $f(\omega^{\prime})=(\exp(\beta \omega^{\prime})+1)^{-1}$
is the Fermi-Dirac function.
Here $\Sigma_{\rm ir}$ is a modified version of $\Sigma$, averaged over the
Fermi surface, but with an extra weighting factor, similar to the
familiar transport factor $1-\cos \theta$.
The actual weighting factor (in the solved electron-phonon
case \cite{Holstein,Allen}) is found from a 
frequency-dependent non-linear
integral equation.  Replacement of the weight factor by 1, turning
$\Sigma_{\rm ir}$ into an ordinary, but $\vec{k}$-averaged, self-energy,
should work fairly well in most cases.
Scher \cite{Scher} did a numerical study which tends to confirm that
the difference between $\Sigma_{\rm ir}$ and $\Sigma$ is small.
Sometimes the anisotropy of $\Sigma$ around the Fermi surface is large.
A modified version of Eq.(\ref{eq:result}) that deals approximately with such cases is presented at the end of the paper.

Eq. (\ref{eq:result}) implies a relation between
$-M(\omega+i\eta)$ and the self-energy which becomes more direct at low frequencies.
Keeping the lowest order (in $\omega$) terms, 
one gets a derivative of the Fermi-Dirac
function $-\partial f(\omega^{\prime})/\partial \omega^{\prime}$
which can be approximated by the
Dirac $\delta(\omega^{\prime})$.
\begin{eqnarray}
\sigma(\omega+i\eta)&\approx& \frac{ine^2}{m} \int_{-\infty}^{\infty} d\omega^{\prime}
\left(-\frac{\partial f(\omega^{\prime})}{\partial \omega^{\prime}} \right)
\nonumber \\
&\times&
\frac{1}{\omega(1-d\Sigma_{{\rm ir},1}(\omega^{\prime})/d\omega^{\prime})
+2i\Sigma_{{\rm ir},2}(\omega^{\prime})} \nonumber \\
& \approx & 
\frac{ine^2/m}{\omega-\omega d\Sigma_{{\rm ir},1}(\omega)/d\omega
+2i\Sigma_{{\rm ir},2}(\omega)}
\label{eq:lowomega}
\end{eqnarray}
Therefore the real part of $-M$ at low frequencies is $\omega
d{\rm Re}\Sigma_{\rm ir}(\omega)/d\omega$, and the imaginary part is 
$-2{\rm Im}\Sigma_{\rm ir}(\omega)$.  If the interesting part of 
${\rm Re}\Sigma$ is odd in $\omega$, and approximately linear, then
$M$ at very low $\omega$ is a lot like $\Sigma$ except for the
factor of 2 in the imaginary part.

In the dc limit, the result $\sigma=ne^2 \tau/m$ is
retrieved, with $1/\tau=-2{\rm Im}\Sigma_{\rm ir}(\omega\rightarrow 0)$.
There are minor differences between this and the more exact 
result found from a
solution of the Boltzmann transport equation.  
These differences arise from $\vec{k}$ dependence,
and disappear when the electron scattering is isotropic.

\section{\label{sec:2} Kubo formula}

The starting point is the Kubo \cite{Kubo} formula for the conductivity.
In an external electric field $\vec{E}(t)=\vec{E}\cos(\omega t)$,
the current operator $j = -e\sum_k v_{kx} c^{\dagger}_k c^{}_k$
acquires an expectation value
$\langle j(t)\rangle={\rm Re}[\sigma(\omega+i\eta)\exp(-i\omega t)]E$,
where the linear response coefficient $\sigma(\omega+i\eta)$ is
\begin{equation}
\sigma(\omega+i\eta)=\frac{i}{\omega}\left[r(\omega+i\eta)+\frac{ne^2}{m}\right]
\label{eq:rtosigma}
\end{equation}
\begin{equation}
r(\omega+i\eta)=i\int_0^{\infty} dt e^{i\omega t-\eta t}\langle [j(t),j(0)]\rangle
\end{equation}
The Hamiltonian ${\cal H}={\cal H}_0 + {\cal H}_1$ has the 
non-interacting part ${\cal H}_0= \sum_k \epsilon_k c^{\dagger}_k c^{}_k$.
The label $k$ is short
for the Bloch wavevector and other quantum numbers $(\vec{k}n\sigma)$.
The state $k$ has energy $\epsilon_k$ and group velocity $\vec{v}_k$.

To obtain a Wick-ordered perturbation theory 
we use an imaginary time ($0\le\sigma\le\beta=1/k_B T$) version
of $r(\omega)$, 
\begin{equation}
r(i\omega_{\mu})=-\int_0^{\beta} d\sigma e^{i\omega_{\mu}\sigma}
\overline{\langle \hat{T} j(\sigma) j(0) \rangle}
\end{equation}
where $j(\sigma)=\exp(\sigma{\cal H})j\exp(-\sigma{\cal H})$.
Angular brackets denote a canonical ensemble temperature
average, and the overbar indicates, if necessary, an average over an ensemble of
randomly distributed impurities.
The Matsubara frequency
$\omega_{\mu}$ is $2\pi\mu/\beta$ and $\mu$ is an integer. 
When analytically continued from 
$i\omega_{\mu}$ to  $\omega + i\eta$ just above the real
$\omega$ axis ($\eta$ is a positive infinitesimal) 
$r(i\omega_{\mu})$ becomes $r(\omega+i\eta)$,
the retarded correlation function needed for the Kubo
formula.

All Feynman graphs for $r(i\omega_{\mu})$ are formally summed
in terms of the exact electron Green's function 
\begin{equation}
G(k,i\omega_{\nu})=\frac{1}{i\omega_{\nu}-\epsilon_k-\Sigma(k,i\omega_{\nu})},
\end{equation}
and the exact vertex function 
$\Gamma(k,k^{\prime},i\omega_{\mu},i\omega_{\nu})$,
where $\omega_{\nu}=2\pi(\nu+1/2)/\beta$.
The exact answer is
\begin{eqnarray}
r(i\omega_{\mu})&=&-\frac{e^2}{\beta}\sum_{k^{}k^{\prime}\nu} v_{k^{\prime}x}
\Gamma(k^{}k^{\prime},i\omega_{\mu},i\omega_{\nu})
\nonumber \\
&\times&G(k^\prime, i\omega_{\nu}+i\omega_{\mu})G(k, i\omega_{\nu})
\label{eq:exact}
\end{eqnarray}
Neither $\Sigma$ nor $\Gamma$ can be calculated exactly.  
A linearized Boltzmann equation is obtained when lowest-order
results for $\Sigma$ and $\Gamma$ are treated consistently.

An explicit formula relating $\sigma$ to $\Sigma$
occurs when $\Gamma$ is replaced by its lowest-order term, 
\begin{equation}
\Gamma(k^{}k^{\prime},i\omega_{\mu},i\omega_{\nu})\rightarrow\Gamma_0=
v_{kx}\delta(k^{},k^{\prime}).
\label{eq:vertex}
\end{equation}
The corresponding answer for $\sigma(\omega+i\eta)$, denoted by $\sigma_0(\omega+i\eta)$,
after continuing to the real frequency axis, and averaging away the k-dependence of $\Sigma(k,\omega+i\eta)$, is
\begin{equation}
\sigma_0(\omega+i\eta)=\frac{ine^2}{m\omega} \int_{-\infty}^{\infty} d\omega^{\prime} 
\frac{f(\omega^{\prime})-f(\omega^{\prime}+\omega)}
{\omega-\Sigma(\omega^{\prime}+\omega+i\eta)
+\Sigma^{\ast}(\omega^{\prime}+i\eta)}.
\label{eq:result0}
\end{equation}
This is the desired approximation, a simplification of the conjectured version, Eq.(\ref{eq:result}).
It is not particularly original.  A derivation for a ``local Fermi liquid'' is given by 
Berthod {\it et al.} \cite{Berthod} It seems worthwhile to 
present a simpler and more general discussion.  A careful derivation
of Eq.(\ref{eq:result0}) is given in the next section.   
This derivation has been available as an arXiv preprint since 2004. \cite{Allen0}  

Unlike the conjectured version Eq. (\ref{eq:result}), the
approximation of Eq. (\ref{eq:vertex}) does not
correctly reproduce the Boltzmann dc conductivity because of the
omission of vertex corrections.  This is related to the fact that
the quasiparticle scattering rate
$1/\tau=-2{\rm Im}\Sigma$ 
differs from the transport scattering rate 
$1/\tau_{\rm tr}=-2{\rm Im}\Sigma_{\rm ir}$ 
by a factor of the type $1-\cos\theta$.
The ``$\cos\theta$'' correction (omitted if the integral equation part of the Boltzmann equation
is neglected) takes into account that small angle $\theta$
scattering events ($\vec{k}\rightarrow\vec{k}^\prime$) do not degrade the current efficiently and make 
smaller contributions to $1/\tau_{\rm tr}$ than to $1/\tau$.
The difference,
except at low temperatures, is likely to be numerically small,
since small angle scattering does not usually play a dominant role.
The version of this applicable to electron-phonon-coupled superconductors was given by Nam. \cite{Nam}

\section{\label{sec:6} Derivation of Eq. (\ref{eq:result0})}

Starting by inserting Eq. (\ref{eq:vertex}) into Eq. (\ref{eq:exact}),
\begin{equation}
r_0(i\omega_{\mu})=-\frac{e^2}{\beta}\sum_k v_{kx}^2
G(k, i\omega_{\nu}+i\omega_{\mu})G(k, i\omega_{\nu})
\label{eq:inexact}
\end{equation}
This approximation, labeled $r_0$, keeps in principle arbitrarily
complicated self-energy graphs in $G$.

The spectral function is defined as
\begin{equation}
G(k, i\omega_{\nu})=\int_{-\infty}^{\infty}
d\omega \frac{A(k,\omega)}{i\omega_{\nu} -\omega}
\label{eq:spectrep}
\end{equation}
\begin{equation}
A(k,\omega)=-\frac{1}{\pi}{\rm Im}G(k,i\omega_{\nu}\rightarrow
\omega+i\eta)
\label{eq:aeqimg}
\end{equation}
where $G(k,\omega+i\eta)$ is the
``retarded'' Green's function, 
\begin{equation}
G(k,\omega+i\eta)=\frac{1}{\omega-\epsilon_k -\Sigma(k,\omega+i\eta)}
\label{eq:gret}
\end{equation}
and $\Sigma(k,\omega+i\eta)=\Delta(k,\omega)-i/2\tau(k,\omega)$
has imaginary part $1/2\tau$ non-negative.  
The approximate correlation function $r_0$ becomes
\begin{eqnarray}
r_0(i\omega_{\mu})&=&-\frac{e^2}{\beta}\sum_k v_{kx}^2
\int_{-\infty}^{\infty}d\omega_1
\int_{-\infty}^{\infty}d\omega_2
A(k,\omega_1)A(k,\omega_2)
\nonumber \\
&\times&\sum_{\nu}\left[\frac{1}{i\omega_{\nu}+i\omega_{\mu} - \omega_1} \
\frac{1}{i\omega_{\nu} - \omega_2}\right]
\end{eqnarray}
The Matsubara sum can be performed exactly,
\begin{equation}
-\frac{1}{\beta}\sum_{\nu}\left[
\frac{1}{i\omega_{\nu}+i\omega_{\mu} - \omega_1} \ 
\frac{1}{i\omega_{\nu} - \omega_2}\right]=\frac{f(\omega_2)-f(\omega_1)}
{\omega_2-\omega_1 +i\omega_{\mu}}.
\end{equation}

The correlation function now is
\begin{eqnarray}
&&r_0(i\omega_{\mu})=-e^2 \int_{-\infty}^{\infty}d\epsilon
\sum_k v_{kx}^2 \delta(\epsilon-\epsilon_k) 
\int_{-\infty}^{\infty}d\omega_1
\nonumber \\ &\times&
\int_{-\infty}^{\infty}d\omega_2
A(k,\omega_1)A(k,\omega_2)
\frac{f(\omega_2)-f(\omega_1)}
{\omega_2-\omega_1 +i\omega_{\mu}},
\label{eq:r0}
\end{eqnarray}
where a gratuitous factor $1=\int d\epsilon \delta(\epsilon-\epsilon_k)$
was inserted.  From Eqs.(\ref{eq:aeqimg},\ref{eq:gret}), the spectral
function has a rapid $\epsilon_k$-dependence,
\begin{equation}
A(k,\omega)=-(1/\pi){\rm Im}[\omega-\epsilon_k -\Sigma(k,\omega+i\eta)]^{-1}.
\label{eq:AA}
\end{equation}
But because of the $\delta$ function in the $k$-sum
\begin{equation}
\sum_k v_{kx}^2 \delta(\epsilon-\epsilon_k)A(k,\omega_1)A(k,\omega_2),
\end{equation}
it is allowed to replace $\epsilon_k$
in the denominators of the spectral functions by $\epsilon$.  
The rapid $\epsilon$-dependence in $A(k,\omega)$ must
be treated carefully, but the remaining weak $k$ dependence of $\Sigma(k,\omega+i\eta)$
in the denominator of $A(k,\omega)$ can often be treated less
carefully.  For many metals, the self-energies $\Sigma(k,\omega+i\eta)$ in the spectral functions 
$-(1/\pi){\rm Im}(\omega-\epsilon -\Sigma)^{-1}$ vary weakly with $\vec{k}$, and can be
replaced by their $k$-average over the Fermi surface,
\begin{equation}
\Sigma(\omega+i\eta)=\sum_k \Sigma(k,\omega+i\eta)\delta(\epsilon_k)
/\sum_k \delta(\epsilon_k).
\label{eq:fsavg}
\end{equation}
In ``conventional'' s-wave superconductors, for example,
anisotropy of the gap function $\Delta(k,\omega)$ is often surprisingly small, and the gap can be
approximated well as $\Delta(\omega)$.  The gap $\Delta(k,\omega)$ is a superconducting 
extension of the normal state self energy.  Serious anisotropy is not forbidden, and is known
to occur in the $T_c$=39K superconductor MgB$_2$ \cite{Choi}, for example.
A modified formula applicable to such cases is given at the end of the paper.
Using Eq.(\ref{eq:fsavg}), the $k$-sum is 
\begin{eqnarray}
\sum_k v_{kx}^2 \delta(\epsilon-\epsilon_k)&=&\frac{1}{\hbar^2}
\sum_k \frac{\partial \epsilon_k}{\partial k_x} \left(-\frac{\partial f}
{\partial k_x}\right)
\nonumber \\ &=&
\frac{1}{\hbar^2} \sum_k \frac{\partial^2 \epsilon_k}{\partial k_x^2}f
=[n/m]_{\rm eff}(\epsilon).
\label{eq:effmass1}
\end{eqnarray}
Here the $\delta$ function was replaced by $-\partial f/\partial \epsilon_k$.
An integration by parts was used to obtain the inverse effective mass
$(\partial^2 \epsilon_k /\partial k_x^2)/\hbar^2$ summed over all
states lower in energy than $\epsilon$.

The range of the remaining $\epsilon$-integration 
is nominally $(-\infty,\infty)$.
However, the factors $A(k,\omega)$ are peaked at
$\epsilon\approx\omega_1$ and $\epsilon\approx\omega_2$.
Thus the integrand is large only if  
$\omega_1$ and $\omega_2$ have similar values.  Because
of the factor $(f(\omega_2)-f(\omega_1))$, they must both lie near
the Fermi energy (one below and one above.)  Therefore the $\epsilon$
integral is dominated by $\epsilon$ near the
Fermi energy.  The value of
$[n/m]_{\rm eff}(\epsilon)$ at the Fermi energy
is $[n/m]_{\rm eff}$, the number of electrons divided
by the effective mass averaged over all states below the
Fermi energy.  An equivalent formula is
\begin{equation}
[n/m]_{\rm eff}=\sum_k v_{kx}^2 \delta(\epsilon_k)=N(0)<v_x^2>
\label{eq:effmass}
\end{equation}

The current correlation function now is
\begin{eqnarray}
r_0(i\omega_{\mu})&=& \left[\frac{n}{m}\right]_{\rm eff} e^2
\int_{-\infty}^{\infty} d\epsilon \int_{-\infty}^{\infty} d\omega_1
\int_{-\infty}^{\infty} d\omega_2
\nonumber \\ &\times&
A(\epsilon,\omega_1)A(\epsilon,\omega_2)
\frac{f(\omega_2)-f(\omega_1)} {\omega_2-\omega_1 +i\omega_{\mu}}.
\label{eq:r1}
\end{eqnarray}
It is necessary to integrate $\epsilon$ carefully over the Lorentzian peaks
of $A(\epsilon,\omega_1)A(\epsilon,\omega_2)$.  Cauchy's theorem
can be used after closing the $\epsilon$-contour by an arc going to
infinity in either the upper or lower half-plane.  The result is
expressed by another identity,
\begin{eqnarray}
\int_{-\infty}^{\infty} d\epsilon
&&\left(\frac{1}{\pi}\right){\rm Im}\left(\frac{1}{\omega_1-\epsilon-\Sigma_1}
\right)
\left(\frac{1}{\pi}\right){\rm Im}\left(\frac{1}{\omega_2-\epsilon-\Sigma_2}
\right)
\nonumber \\
&=&-\left(\frac{1}{\pi}\right){\rm Im}\left(\frac{1}{\omega_1-\omega_2
-\Sigma_1+\Sigma_2^{\ast}}\right).
\label{eq:identity2}
\end{eqnarray}
The proof is elementary but tedious.  The current correlation function is now
\begin{eqnarray}
r_0(i\omega_{\mu})&=& \left[\frac{n}{m}\right]_{\rm eff} \frac{e^2}{\pi}
\int_{-\infty}^{\infty} d\omega_1 \int_{-\infty}^{\infty} d\omega_2
\frac{f(\omega_2)-f(\omega_1)} {\omega_2-\omega_1 +i\omega_{\mu}}
\nonumber \\ &\times&
{\rm Im}\left(\frac{1}{\omega_1-\omega_2 -\Sigma_1+\Sigma_2^{\ast}}\right).
\end{eqnarray}

The function
$r_0(\omega+i\eta)$ is now just $r_0(i\omega_{\mu})$ with
$i\omega_{\mu}$ replaced by $\omega+i\eta$. 
The only complex
quantity in the formula for $r_0(\omega+i\eta)$ is the factor
$(\omega_2-\omega_1 + \omega +i\eta)^{-1}$, 
so the real part ${\rm Re}\sigma_0(\omega+i\eta)=
{\rm Im}r_0(\omega+i\eta)/\omega$ (Eq.(\ref{eq:rtosigma})) is
\begin{eqnarray}
{\rm Re}\sigma_0(\omega+i\eta)&=&\left[\frac{n}{m}\right]_{\rm eff} \frac{e^2}{\omega}
\int_{-\infty}^{\infty} d\omega_1 \int_{-\infty}^{\infty} d\omega_2
[f(\omega_2)-f(\omega_1)]
\nonumber \\ &\times&
\delta(\omega_2-\omega_1 + \omega)
{\rm Re}\left(\frac{i}{\omega_1-\omega_2 -\Sigma_1+\Sigma_2^{\ast}}\right)
\nonumber \\
&=&\left[\frac{n}{m}\right]_{\rm eff}e^2
\int_{-\infty}^{\infty} d\omega^{\prime}
\left[\frac{f(\omega^{\prime})-f(\omega^{\prime}+\omega)}{\omega}\right]
\nonumber \\ &\times&
{\rm Re}\left(\frac{i}{\omega -\Sigma(\omega^{\prime}+\omega+i\eta)
+\Sigma^{\ast}(\omega^{\prime}+i\eta)}\right).
\label{eq:sigma1}
\end{eqnarray}
The function $\sigma_0(\omega+i\eta)$ is specified
by the requirements of being analytic for ${\rm Im}\omega>0$, vanishing
sufficiently rapidly as $\omega\rightarrow\infty$, and agreeing
with Eq. (\ref{eq:sigma1}).  It is necessary and sufficient to 
remove the ``real part'' designator from both sides.  This is the 
derivation of Eq. (\ref{eq:result0}).

Here is a modification of Eq.(\ref{eq:result0}) that doesn't ignore anisotropy of $\Sigma(k,\omega+i\eta)$ as $\vec{k}$
varies around the Fermi surface.
\begin{eqnarray}
&&\sigma_0(\omega+i\eta)=\frac{ie^2}{\omega} \int_{-\infty}^{\infty} d\omega^{\prime} 
[f(\omega^{\prime})-f(\omega^{\prime}+\omega)] \nonumber \\
&\times&\sum_k \frac{v_{kx}^2 \delta(\epsilon_k - \epsilon_F)}
{\omega-\Sigma(k,\omega^{\prime}+\omega+i\eta)
+\Sigma^{\ast}(k,\omega^{\prime}+i\eta)} 
\label{eq:anis}
\end{eqnarray}
To derive this, go back to Eq.(\ref{eq:r0}) but do not use the isotropic form Eq.(\ref{eq:fsavg})
for $\Sigma$, and do not make use of Eq.(\ref{eq:effmass1}).  The factor $[n/m]_{\rm eff}$
no longer appears outside the integrals, but $\sum_k v_{kx}^2 \delta(\epsilon_k - \epsilon)$
appears inside the $d\epsilon$ integral in Eq.(\ref{eq:r1}).  It is no longer
possible to use Eq.(\ref{eq:identity2}), unless an approximation is made, namely that the
$\vec{k}$-dependence of $v_{kx}^2$ and of $\Sigma(k,\omega+i\eta)$ is not too rapid.  There can
be a large variation of both  $v_{kx}^2$ and $\Sigma(k,\omega+i\eta)$ as $\vec{k}$ moves around the 
Fermi surface.  However, the variation perpendicular
to the Fermi surface, as $\epsilon_k$ changes on the scale of the relevant
infrared $\omega$'s, must be small.
Then one can ignore the $\epsilon$-dependence of both 
$v_{kx}^2$ and $\Sigma(k,\omega+i\eta)$, and recover the use of Eq.(\ref{eq:identity2}).
Then Eq.(\ref{eq:anis}) follows.  A version of this approximation was used by Hussey {\it et al.} \cite{Hussey}
for analysis of the normal state of cuprates.

\section{Acknowledgements}
This work was supported in part by DOE grant no. DE-FG02-08ER46550.


\begin{thebibliography}{99}

\bibitem{Holstein1} T. Holstein, 
{\it Optical and infrared volume absorptivity of metals},
Phys. Rev. {\bf 96}, 535 (1954).

\bibitem{Joyce} R. R. Joyce and P. L. Richards, 
{\it Phonon contribution to the far-infrared absorptivity of superconducting and normal lead},
Phys. Rev. Lett. {\bf 24}, 1007 (1970).

\bibitem{Farnworth} B. Farnworth and T. Timusk,
{\it Phonon density of states of superconducting lead},
Phys. Rev. B {\bf 14}, 5119 (1976).

\bibitem{Gotze} W. G\"otze and P. W\"olfle, 
{\it Homogeneous dynamical conductivity of metals}, 
Phys. Rev. B {\bf 6}, 1226 (1972).

\bibitem{Timusk} J. Huang, T. Timusk, and G. D. Gu,
{\it High transition temperature superconductivity in the absence of the 
magnetic resonance mode,} 
Nature {\bf 427}, 714 (2004).

\bibitem{Holstein} T. Holstein, 
{\it Theory of transport phenomena in an electron-phonon gas},
Ann. Phys. (N.Y.) {\bf 29}, 410 (1964).

\bibitem{Allen} P. B. Allen, 
{\it Electron-phonon effects in the infrared properties of metals}, 
Phys. Rev. B {\bf 3}, 305 (1971).

\bibitem{Scher} H. Scher, 
{\it Far-infrared absorptivity of normal lead},
Phys. Rev. Letters {\bf 25}, 759 (1970).

\bibitem{Kubo} R. Kubo, 
{\it Statistical-mechanical theory of irreversible processes. 1. general theory and simple applications to
magnetic and conduction problems}, 
J. Phys. Soc. Jpn. {\bf 12}, 570 (1957).

\bibitem{Berthod}  C. Berthod, J. Mravlje, X. Deng, R. Zitko, D. van der Marel, and A. Georges,
{\it Non-Drude universal scaling laws for the optical response of local Fermi liquids},
Phys. Rev. B {\bf 87}, 115109 (2013).

\bibitem{Allen0} P. B. Allen, 
{\it Electron self-energy and generalized Drude formula for infrared conductivity of metals},
arXiv:cond-mat/0407777.

\bibitem{Nam} S. B. Nam,
{\it Theory of Electromagnetic Properties of Superconducting and Normal Systems. I},
Phys. Rev. {\bf 156}, 470 (1967).

\bibitem{Choi} H. J. Choi, D. Roundy, H. Sun, M. L. Cohen, and S. G. Louie,
{\it First-principles calculation of the superconducting transition in MgB$_2$ within the anisotropic Eliashberg formalism},
Phys. Rev. B {\bf 66}, 020513 (2002).

\bibitem{Hussey} N. E. Hussey, J. C. Alexander, and R. A. Cooper,
{\it Optical response of high-Tc cuprates: Possible role of scattering rate saturation and in-plane anisotropy},
Phys. Rev. B {\bf 74}, 214508 (2006).


\end{thebibliography}
\end{document}